\pdfoutput=1
\documentclass[sigconf,authorversion]{acmart}
\usepackage{booktabs}
\usepackage{listings}
\usepackage{cleveref}

\setcopyright{none}

\acmDOI{}

\acmConference[K-CAP]{the Ninth International Conference on Knowledge Capture}{Dec. 4 - 6, 2017}{Austin, TX, USA} 
\acmYear{2017}

\begin{document}
\title[Content Recommendation through Semantic Annotation of User Reviews and Linked Data]{Content Recommendation through Semantic Annotation of User Reviews and Linked Data -- An Extended Technical Report}

\author{Iacopo Vagliano}
\affiliation{
  \institution{ZBW - Leibniz Information Centre for Economics}
  \city{Kiel}
  \state{Germany}
}
\email{i.vagliano@zbw.eu}

\author{Diego Monti}
\affiliation{
  \institution{Politecnico di Torino}
  \city{Turin}
  \state{Italy}
}
\email{diego.monti@polito.it}

\author{Ansgar Scherp}
\affiliation{
  \institution{ZBW - Leibniz Information Centre for Economics}
  \city{Kiel}
  \state{Germany}
}
\email{a.scherp@zbw.eu}

\author{Maurizio Morisio}
\affiliation{
  \institution{Politecnico di Torino}
  \city{Turin}
  \state{Italy}
}
\email{maurizio.morisio@polito.it}

\begin{abstract}
Nowadays, most recommender systems exploit user-provided ratings to infer their preferences. However, the growing popularity of social and e-commerce websites has encouraged users to also share comments and opinions through textual reviews. In this paper, we introduce a new recommendation approach which exploits the semantic annotation of user reviews to extract useful and non-trivial information about the items to recommend. It also relies on the knowledge freely available in the Web of Data, notably in DBpedia and Wikidata, to discover other resources connected with the annotated entities. We evaluated our approach in three domains, using both DBpedia and Wikidata. The results showed that our solution provides a better ranking than another recommendation method based on the Web of Data, while it improves in novelty with respect to traditional techniques based on ratings. Additionally, our method achieved a better performance with Wikidata than DBpedia.
\end{abstract}

%
%


\maketitle

\section{Introduction}\label{sec:intro}
The Web has evolved from an information space to share textual documents into a medium to distribute structured data. 
Linked Data\footnote{\url{http://linkeddata.org}} is a set of best practices for publishing and interlinking data on the Web and it is the base of the Web of Data, an interconnected global knowledge graph. Because of the increased amount of machine-readable knowledge freely available on the Web, there is a high interest in investigating how such information can be used to improve recommender systems.

Currently, most recommender systems exploit ratings to infer user preferences, although the growing popularity of social and e-commerce websites has encouraged users to write reviews. These reviews enable recommender systems to represent the multi-faceted nature of users' opinions and build a fine-grained preference model, which cannot be obtained from overall ratings~\cite{Chen2015}. Additionally, recommender systems may take advantage of reviews because they are harder to fake than ratings, are richer of information, and users may struggle to express their preference as ratings. Some studies have also documented the positive influence of product reviews on the decision processes of new users~\cite{chatterjee2001online, Kim:2007:ISI:1282100.1282157}.

We address the issue of mining reviews and show how the extracted information, combined with Linked Data, can be exploited in recommendation tasks. On one side Linked Data can provide a rich representation of the items to be recommended since they include interesting features. For example, movies represented in DBpedia\footnote{\url{http://dbpedia.org}} contain classical information such as cast and director, but also some unexpected relations, e.\,g. both \emph{Braveheart} and \emph{Saving Private Ryan} won the Best Sound Editing Academy Award. On the other side, reviews may reveal additional connections among items.
For instance, various reviews of \emph{Interstellar} mention Stanley Kubrick, although in DBpedia there is not a direct link between these two resources.
 
We propose a new recommendation approach that semantically annotates reviews to extract useful information from them. The annotated entities and the knowledge freely available in the Web of Data are then combined to discover additional resources and generate recommendations. Our method can exploit any dataset available in the Web of Data to provide recommendations, although we rely on DBpedia and Wikidata\footnote{\url{https://www.wikidata.org}} in our implementation. 

More precisely, we conducted an offline study to find the best configuration of our technique for these two datasets and comparatively evaluate our approach against a Linked Data based and some traditional algorithms based on ratings. We performed the study in the movie, book, and music domains, and the evaluation took into account different properties of recommender systems, i.\,e. prediction accuracy (both in terms of ratings and ranking), diversity, and novelty.
In fact, not only accuracy is important: recommendations all known to users or all of the same kind (e.\,g., all movies already watched or all movies of the same genre or with the same actor) may not satisfy them, although they match their taste (users expect to discover new movies to watch and may be bored of dramas, although they generally like it). The results showed that our method achieved the highest diversity, provided a better accuracy than the method based on Linked Data, and increased the novelty of recommendations with respect to traditional techniques.

The contribution of this paper is threefold. Firstly, we exploit state-of-the-art semantic annotation techniques to extract, from user reviews, useful and non-trivial information about the items to recommend. The extracted entities are resources in the Web of Data; thus we can discover additional knowledge through their links. Secondly, we rely on the annotated and discovered entities to provide recommendations, taking into account their occurrence in the reviews and their relationships in the Web of Data. Thirdly, we validate our approach by evaluating its effectiveness through an offline study conducted in the movie, book, and music domains.

The remainder of this paper is organized as follows: Section~\ref{sec:soa} reviews related works; Section~\ref{sec:approach} presents our approach; Section~\ref{sec:eval} describes the evaluation method, while Section~\ref{sec:res} shows the obtained results and Section~\ref{sec:discussion} discusses them; Section~\ref{sec:concl} provides the conclusions.

\section{Related Work}\label{sec:soa}
In this section, we distinguish among works that exploit user reviews for recommendation tasks and studies which discuss Linked Data based recommender systems. In the best of our knowledge, we are the first to combine the use of reviews and Linked Data. 

\paragraph*{Review-based Recommender Systems}
The exploitation of user reviews in recommender systems is a well-known research topic. Some techniques try to tackle the problem of building the profile of users by analyzing their reviews, while others focus on the identification of the main features of the items to recommend, as Cheng et al.~\cite{Chen2015} summarized in their survey.
Different strategies have been proposed in the literature to address the latter problem. Some researchers have suggested methods able to identify the sentiment associated with the features of an item exploiting a domain-specific ontology~\cite{aciar_informed_2007} or its technical description~\cite{yates_shopsmart_2008}. A common aspect of these techniques is that the possible features are already available before performing the analysis.
However, there are also approaches for unsupervised extraction of product features and sentiment from reviews~\cite{DBLP:journals/coling/QiuLBC11, DBLP:conf/fskd/SomprasertsriL10}. Since we use Knowledge Graphs to extract and expand features from the reviews, we do not apply those unsupervised extraction techniques. Nevertheless, it may be interesting in the future to combine both approaches and first conduct an unsupervised extraction of item features and sentiment from reviews and subsequently perform an expansion via knowledge graphs.

Another possibility is to identify the main characteristics of an item with the help of natural language processing methods, without any previous knowledge of the context. For example, a popular technique considers bigrams that frequently occur in reviews and that are associated with a word expressing an emotion~\cite{dong_opinionated_2013}. In this case, the goal of the recommender system is suggesting items with the same features of the ones liked by the target user, but with a better global sentiment.
In the best of our knowledge, there is only one attempt to exploit user reviews for recommendation tasks using semantic annotation. Dzikowski et al.~\cite{Dzikowski2012} applied semantic annotation to reviews while users are editing them. Their goal was to produce annotated reviews of restaurants through Linked Data in order to generate tags to be associated with the reviewed items. In contrast, we apply semantic annotation and find related items.

\paragraph*{Linked Data based Recommender Systems}\label{sec:ld}
In the past, some studies reviewed different Linked Data based recommender systems that were proposed in the literature~\cite{DiNoia2015, CPE:CPE3449}. Typically these recommender systems consider the relationships among resources by taking into account the existing links in the Web of Data and use these relationships to measure the semantic similarity of the resources. Such relationships can be direct links or paths between the items to recommend. In the following, we summarize the main works, although none of these exploit reviews. 
Damljanovic et al.~\cite{LDBCR} suggested domain experts in an open innovation scenario. Their approach generates recommendations by discovering related resources through hierarchical or transversal relationships in DBpedia. Passant~\cite{Passant2010} presented \emph{dbrec}, a music recommender system, which mainly relies on a measure named Linked Data Semantic Distance (LDSD). This measure is based on the number of direct and indirect links between two resources. Heitmann and Hayes~\cite{Heitmann_c.:using} also proposed a recommender system which exploited Linked Data to mitigate the new-user, new-item and sparsity problems of collaborative recommender systems.
More recently, Musto et al.~\cite{Musto:2016:SGR:2930238.2930249} studied the impact of the knowledge available in the Web of Data on the overall performance of a graph-based recommendation algorithm. 
Vagliano et al.~\cite{ReDyAl} presented a recommendation algorithm based on Linked Data which exploits existing relationships between resources by dynamically analyzing both their categories and their explicit references to other resources. 
Di Noia et al. described a model-based approach to provide content-based recommendations with Linked Data~\cite{DiNoia:2012:EWD:2365952.2366007}. 
Ostuni et al.~\cite{Ostuni2014} defined a neighborhood-based graph kernel for matching graph-based item representations. 
Di Noia et al.~\cite{noia_sprank_2016} introduced SPrank, a hybrid algorithm which extracts semantic path-based features from DBpedia and computes recommendations using Learning to Rank.

\section{Approach}\label{sec:approach}
\begin{figure}
\centering
\includegraphics[width=0.4\textwidth]{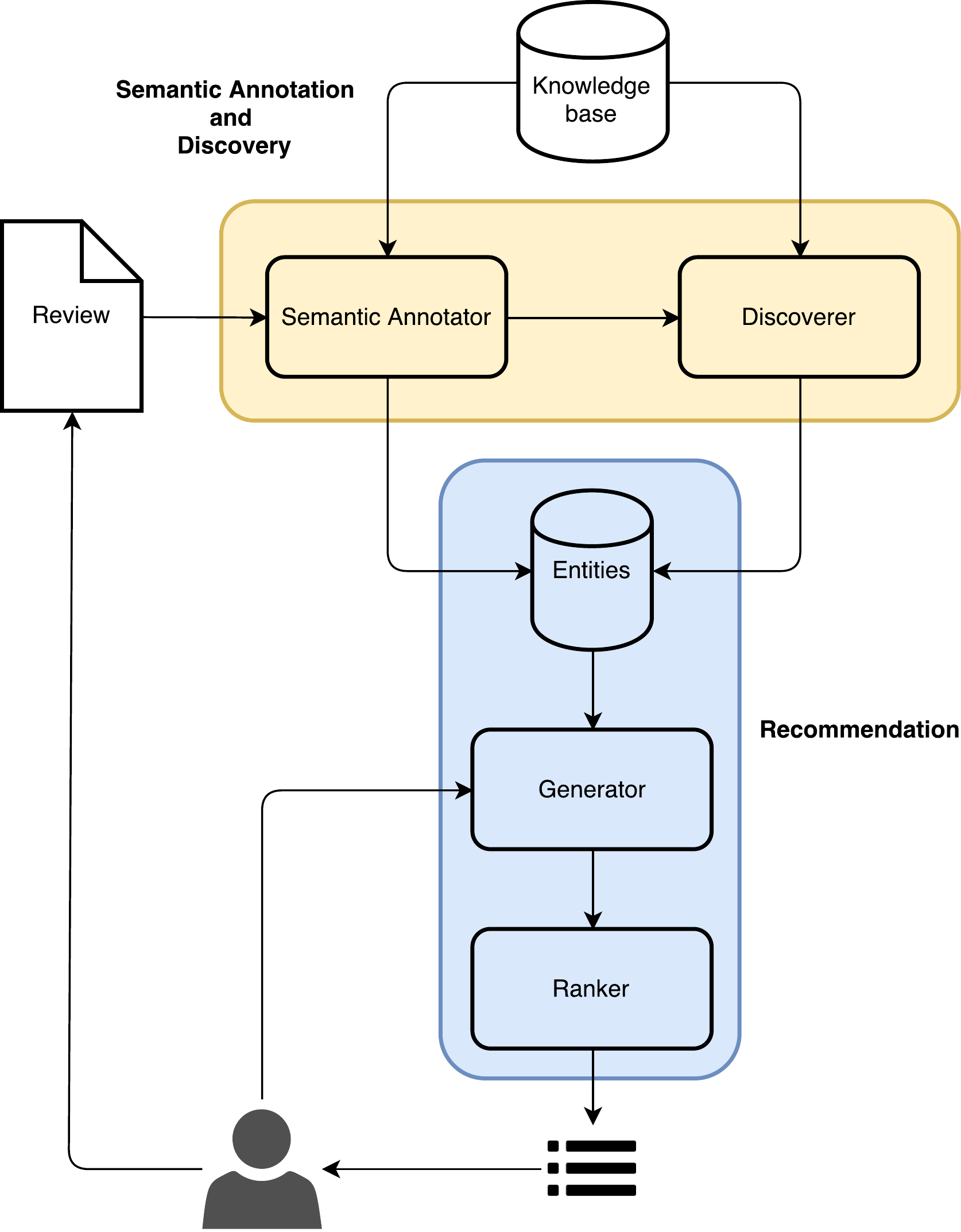}
\caption{SemRevRec architecture}
\label{fig:arch}
\end{figure}

The architecture of SemRevRec is depicted in Figure~\ref{fig:arch}. The system consists of two main modules which are highlighted with different colors: semantic annotation and discovery, and recommendation. The former is responsible for feeding the recommender system with semantically annotated entities and Linked Data through the knowledge base, while the latter provides recommendations to users. Every time a new review is submitted, the system executes the semantic annotation and discovery steps and possibly adds new entities, while the recommendation process can start when the user provides an initial item. The recommendation module works online, while the semantic annotation and discovery are done offline. Initially, some reviews are annotated and the resulting entities are used to discover additional entities through Linked Data. Each of these two modules is made up of the submodules depicted, which are responsible for specific steps of the whole process: annotation, discovery, generation of recommendations, and their ranking. The storage of entities is not a step, but the corresponding database is a transversal submodule used by all the others. 

SemRevRec deals with the annotated or discovered entities and the items to recommend. We consider the items a particular type of entities since SemRevRec recommends items which may be annotated or discovered entities, although an item may not appear as an entity in the system, e.\,g., a movie is reviewed but was never annotated or discovered. However, this does not mean that an entity corresponding to such film does not exist in the considered knowledge base. Semantic annotation and discovery are explained in Section~\ref{sec:annotation}, while recommendation is presented in Section~\ref{sec:recommendation}.

Although our approach is not bounded to a particular domain or knowledge base available in the Web of Data, in our implementation, we focus on movies, books, and music, while we rely on DBpedia and Wikidata to identify possible differences between these two knowledge bases. We chose them for annotation and discovery because they are two of the main datasets in the Web of Data, and have a vast amount of resources represented which belongs to a variety of domains. We used reviews from IMDb\footnote{\url{http://www.imdb.com}} for movies, LibraryThing\footnote{\url{https://www.librarything.com}} for books, and Amazon\footnote{\url{https://www.amazon.com}} for music.

\subsection{Semantic Annotation and Discovery}\label{sec:annotation}
Semantic annotation is the process of annotating textual or multimedia contents with semantic tags to add information about their meaning~\cite{Saathoff2010}. In written text, this can be done by associating a URI to the recognized entities. We considered two popular semantic annotators that rely on Wikipedia: AIDA~\cite{aida_2011} and DBpedia Spotlight~\cite{spotlight_2013}. They are both capable of disambiguating entities according to the surrounding context: this is useful because users frequently write acronyms and abbreviations. We finally selected AIDA because it is more accurate according to an independent comparison~\cite{gangemi_comparison_2013}.

The module of semantic annotation and discovery analyzes the text of the reviews and stores the identified entities in a relational database. The URI of each annotated entity is associated with the URI of the reviewed item and with the occurrence of that entity in all the reviews of that item. In effect, the same entity may appear again in reviews regarding another item. AIDA is capable of identifying and disambiguating the entities mentioned in the review considering, by default, the ones available in YAGO\footnote{\url{http://www.yago-knowledge.org}}.

The AIDA resources are mapped with the equivalent ones available in DBpedia exploiting the similar structure of the URIs. For example, \texttt{yago-res:The\_Matrix} corresponds to \texttt{dbr:The\_Matrix} because their URIs where generated starting from the title of the same Wikipedia article. In contrast, the mapping between DBpedia and Wikidata relies on the \texttt{owl:sameAs} predicate available in DBpedia. If the same entity corresponds to more than one in the other knowledge base, it is ignored in order to avoid probable inconsistencies. The same holds if there is no \texttt{owl:sameAs} property, although DBpedia is well linked to Wikidata. In principle, it is also possible to perform the semantic annotation phase relying on a custom knowledge base, but AIDA is provided with a precomputed database that includes all the necessary information for annotating with YAGO. In our case, since DBpedia and Wikidata are both well interlinked with YAGO, it was less time consuming computing the mapping than the information needed by the annotator for these two knowledge bases.

Finally, the types of each entity are obtained from the target knowledge base, optionally considering only a subset of them (e.\,g. only the DBpedia ontology types, such as \texttt{dbo:Film}). This is done in order to minimize the amount of information retrieved and to reduce the time required for this operation. The types are stored locally because they are not expected to change often and reading them from a relational database is more efficient than querying the original knowledge base.

Semantic annotation allows SemRevRec to exploit Linked Data for retrieving additional entities. This is possible because the annotated entities are also resources in the Web of Data. Thus, the discoverer can find resources which are related to the annotated entities in order to enable our system to recommend more items. Reviews are a source of non-trivial relations: for example, in a movie recommendation scenario, a user can mention a movie which reminds him the reviewed one because of the colors, the setting, or the atmosphere, and these features are hardly available as Linked Data. At the same time, Linked Data can enrich information coming from users. For instance, they enable the discoverer to obtain other movies in which an actor mentioned in a review played. In order to do so, the discovery can take into account various properties, from more traditional, such as the genre, the director, or the actors of the movie reviewed, to more unexpected ones, such as other movies shot in the same country.

Given the annotated entities, the discoverer retrieves from the knowledge base other relevant entities through SPARQL queries. It relies on some properties which can be configured and depend on the domain and on the dataset considered. The discovery is not bounded to a particular knowledge base or domain. On the contrary, this approach is fairly general since it relies only on RDF and SPARQL. In our implementation, we considered DBpedia and Wikidata, and we focused on movie, book, and music recommendations. Table~\ref{table:disc} summarizes the properties that we selected for discovering further items to recommend starting from the entities available in the reviews.

\begin{table}
\centering
\caption{Properties considered for discovery}
\label{table:disc}
\begin{tabular}{@{}lll@{}}
\toprule
Domain & DBpedia               & Wikidata          \\ \midrule
Movie  & \texttt{dbo:starring} & \texttt{wdt:P161} \\
Movie  & \texttt{dbo:director} & \texttt{wdt:P57}  \\
Book   & \texttt{dbo:author}   & \texttt{wdt:P50}  \\
Music  & \texttt{dbo:artist}   & \texttt{wdt:P175} \\
Music  & \texttt{dbo:writer}   & \texttt{wdt:P676} \\ \bottomrule
\end{tabular}
\end{table}

More specifically, the discoverer reads the annotated entities stored during the semantic annotation phase. The discoverer is able to obtain all the resources which have the given entities as an object of the selected properties.
For example, in the movie domain, we selected \texttt{dbo:starring} and \texttt{dbo:director} in the case of DBpedia because most of the annotated properties, when not movies, were actors and directors. This allows the system to discover other movies from the same director or actor named in a given review. Sometimes directors or actors not involved in the movie were also mentioned for comparison. The discoverer can retrieve other movies from these entities, which are relevant for the user who wrote the review, thus can also be of interest for other users. Similarly to movies, we selected \texttt{dbo:author} for books as well as \texttt{dbo:artist} and \texttt{dbo:writer} for music because most of the annotated entities were authors, artists or writers when not books and songs, respectively. It is possible to exploit both direct and inverse properties in the discovery.
 
The discoverer stores the discovered entities in a relational database for efficiency reasons. The URI of each discovered entity is associated with the URI of the annotated entity through which it was discovered, and, optionally, with the LDSD measure~\cite{Passant2010} between them. This measure is inversely proportional to the number of links between two resources: more links result in a lower distance. Each discovered entity may be found through more than a single annotated entity. The LDSD can be exploited in the ranking phase, which is described in Section~\ref{sec:ranking}. However, since its computation is expensive due to the various SPARQL queries involved, it may be optionally skipped to speed up the discovery step. Obviously, in this case, the LDSD measure does not contribute to the ranking.

\subsection{Recommendation}\label{sec:recommendation}
The recommendation process consists of two main steps: the generation of the candidate recommendations and their ranking. Given an initial item, SemRevRec retrieves all the entities which are related to the initial item and then ranks them.

Firstly, the system selects the annotated entities which were mentioned in the reviews of the initial item. Afterwards, it obtains the entities which mention the initial item, i.\,e., entities whose reviews generated an annotated entity that corresponds to the initial item. For example, if the initial item is \emph{Interstellar} and a review of \emph{2001: A Space Odyssey} mention \emph{Interstellar}, then \emph{2001: A Space Odyssey} is considered as a candidate recommendation. Then, SemRevRec optionally retrieves the discovered entities. They may include entities discovered through the initial item.
For instance, if the initial item is \emph{Interstellar} and \emph{The Dark Knight} was previously discovered because both these movies have been directed by Christopher Nolan, \emph{The Dark Knight} is selected. The same holds if \emph{Interstellar} was discovered from \emph{The Dark Knight}, i.\,e., Christopher Nolan was annotated in the reviews of the latter. Similarly, the entities discovered through other entities which were annotated in the reviews of the initial item are relevant. For example, if \emph{Interstellar} is the initial item, Stanley Kubrick was annotated in one of its reviews, and \emph{2001: A Space Odyssey} was discovered through Stanley Kubrick, then \emph{2001: A Space Odyssey} is a candidate recommendation. It is possible to configure the generator to include in the candidate recommendations the discovered entities or not. It is also possible to specify the minimum occurrence required for entities to be included in the candidate recommendation set, which is expressed as a percentage of the maximum occurrence of entities in the reviews of the item considered.

\subsection{Ranking Functions}\label{sec:ranking}
Finally, SemRevRec ranks the candidate recommendations. We defined three different ranking functions. The first is presented in Equation~\ref{eq:r1} and takes into account only the occurrence $occur(i)$ of the entities available in the reviews. $occur(i)$ is equal to the number of reviews of an initial item $i_{in}$ where an entity $i$ is annotated, plus the number of reviews of $i$ where $i_{in}$ is annotated (if any). However, the entity $i$ can be annotated or discovered. For the latter, the occurrence of the entity through which it was discovered is used. The $\alpha$ coefficient is $1$ if $i$ is an annotated entity. Otherwise, it can be configured to a custom value (the default is $0.5$) to weight the contribution of a discovered entity to the ranking. To obtain a value between $0$ and $1$, R1 is normalized to the maximum occurrence of entities $j$ which belong to the candidate recommendation set $CR$.
\begin{equation}\label{eq:r1}
\mathit{R1}(i) = \frac{\alpha \cdot \mathit{occur(i, i_{in})}}{max_{j \in \mathit{CR}}(\mathit{occur(j, i_{in})})}
\end{equation}

The second ranking function (Equation~\ref{eq:r2}) also considers the LDSD measure between each discovered entity and the entity through which it was discovered. This avoids assigning the same value to all the entities discovered through the same annotated entity as R1 does. 
As for R1, the entity $i$ can be annotated or discovered. The $\beta$ coefficient is $1$ if $i$ is an annotated entity, $0.5$ otherwise. The $\gamma$ coefficient is $0.5$ for discovered entities, $0$ otherwise. In this way, R2 returns a number between $0$ and $1$, which is equal to R1 for the annotated entities, while, for the discovered entities, it is the average of R1 and $LDSD(i,i_o)$, where $i_o$ is the entity through which $i$ was discovered.
\begin{equation}\label{eq:r2}
\mathit{R2}(i) = \beta \cdot \mathit{R1}(i) + \gamma \cdot (1 - \mathit{LDSD}(i, i_o))
\end{equation}

The third ranking function (Equation~\ref{eq:r3}) considers the LDSD measure between an entity $i$ and the initial item $i_{in}$. The coefficients $\eta$ and $\kappa$ can be set to custom values and they allow the ranker to weight differently the contribution of the occurrence in the review (given by R2) and Linked Data (through the LDSD measure).
\begin{equation}\label{eq:r3}
\mathit{R3}(i) = \eta \cdot \mathit{R2}(i) + \kappa \cdot (1- \mathit{LDSD}(i, i_{in}))
\end{equation}

LDSD measures between discovered entities and the entities through which they were discovered need to be precomputed at discovery time (see Section \ref{sec:annotation}) to enable SemRevRec to exploit R2, LDSD measures between entities in $CR$ and the initial item need to be computed while ranking. In the latter case, the ranking time is increased.
\section{Evaluation Procedure}\label{sec:eval}
We evaluated the performance of SemRevRec with two offline experiments conducted in the movie, book, and music domains. The purpose of the first experiment is to understand the impact of the ranking function, the discovery, the occurrence threshold, and the coefficients of R3. Furthermore, we performed the first experiment two times, first relying on DBpedia and then on Wikidata, to assess the effect of the exploited knowledge base on the quality of the recommended items. The aim of the second experiment is to compare our proposal with traditional recommendation techniques that rely on ratings and a recommender system based on Linked Data.

In order to conduct both experiments, we obtained from IMDb, LibraryThing, and Amazon the user reviews regarding all the items included in the MovieLens~1M\footnote{\url{http://grouplens.org/datasets/movielens/1m/}}, the LibraryThing\footnote{\url{http://www.macle.nl/tud/LT/}} and the HotRec 2011 LastFM\footnote{\url{http://ir.ii.uam.es/hetrec2011/datasets/lastfm/readme.txt}} datasets of user ratings.

The items of such rating datasets were mapped with the corresponding entities available in DBpedia relying on the work of Di Noia et al.~\cite{noia_sprank_2016}. Moreover, their equivalent entities in Wikidata were obtained from DBpedia itself, as described in Section~\ref{sec:annotation}. For the purpose of retrieving the user reviews, Wikidata was exploited in order to discover the IMDb identifiers of the movies available in the MovieLens~1M dataset. On the contrary, the LibraryThing dataset already contained the references useful for obtaining the reviews. Regarding the musical artists present in the HotRec 2011 LastFM dataset, we relied on the search feature of Amazon for identifying their most reviewed musical work.

\begin{table}
\centering
\caption{Datasets and reviews statistics}
\label{table:stats}
\begin{tabular}{@{}lrrr@{}}
\toprule
                   & Movie     & Book         & Music   \\ \midrule
Users              & 6,040     & 7,279        & 1,892   \\
Items              & 3,706     & 37,232       & 17,632  \\
Ratings            & 1,000,209 & 2,056,487    & 92,834  \\
Reviews            & 559,858   & 363,791      & 669,978 \\
Distinct entities  & 107,468   & 77,120       & 70,762  \\
Total entities     & 574,435   & 303,705      & 296,777 \\ \bottomrule
\end{tabular}
\end{table}

\begin{figure}
\centering
\includegraphics[width=0.45\textwidth]{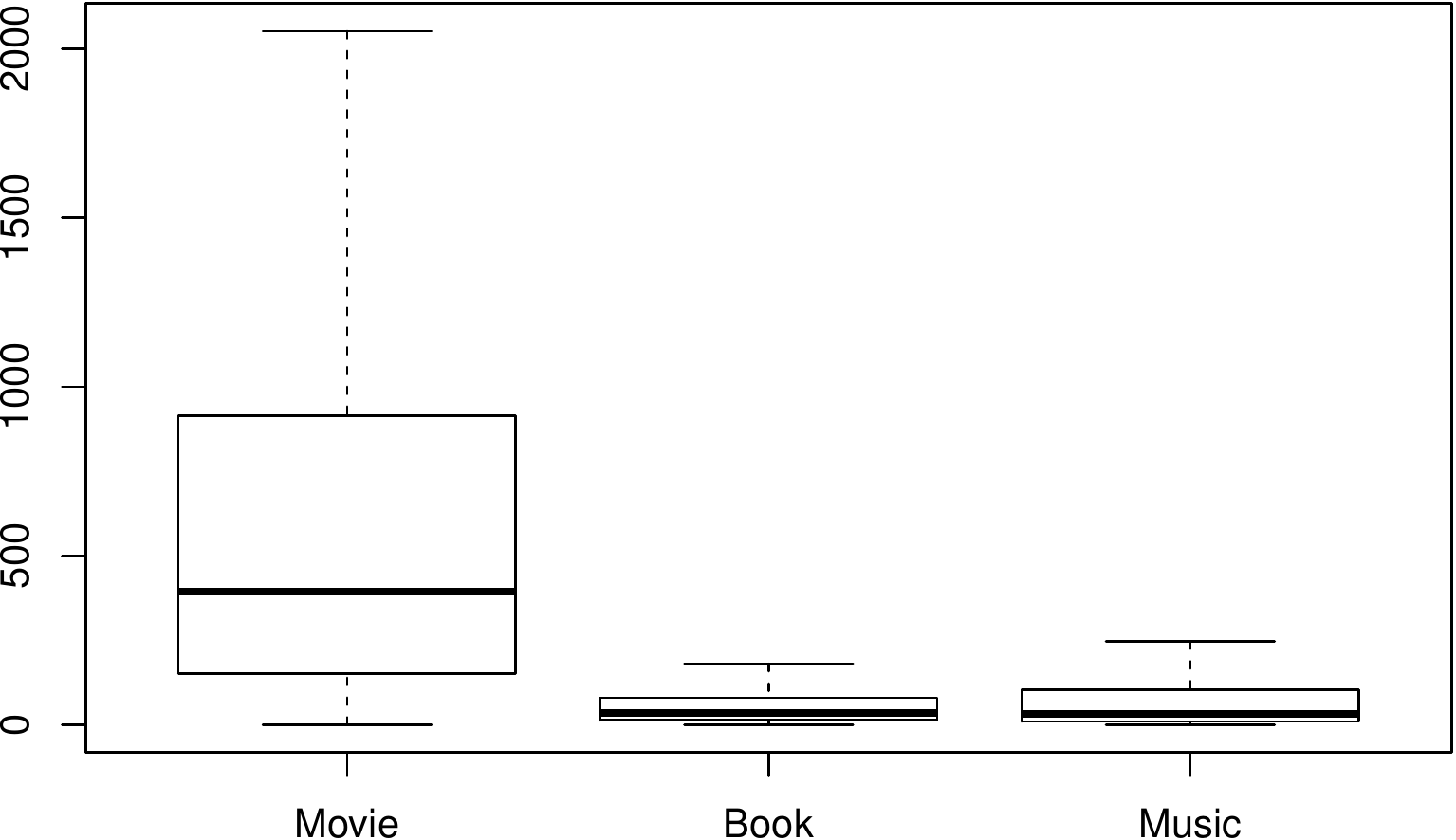}
\caption{Distribution of entities extracted from the reviews of the items per domain}
\label{fig:boxplot}
\end{figure}

Table~\ref{table:stats} lists several statistics regarding the exploited rating datasets and the analyzed reviews in the three domains considered. It is worth noting that the HotRec 2011 LastFM dataset contains a limited number of ratings with respect to the other datasets and, for this reason, it is the most sparse one. The LibraryThing dataset includes a considerable number of items, even if fewer reviews are available in the book domain. Regarding the outcome of the semantic annotation, the number of distinct and total entities identified in user reviews is reported. The ratio between these two values may be considered a measure of the variety of the mentioned topics. According to this measure, the reviews about movies are the most varied ones in terms of entities.

Figure~\ref{fig:boxplot} displays the boxplots representing the distributions of the number of annotated entities per each item according to the domain, excluding the outliers for graphical reasons. Given the interquartile range $IQR = Q3 - Q1$, all data points not belonging to the interval $(Q1 - 1.5 \cdot IQR; Q3 + 1.5 \cdot IQR)$ are considered outliers. It is clear that movie reviews are fairly different from the other ones. This may be related to the higher ratio between reviews and items in the movie domain.

In order to perform the evaluations, a 5-fold cross-validation was executed.
Here, we considered ratings as positive if their score was greater than 3 on a scale from 1 to 5 for MovieLens, greater than 6 on a scale from 1 to 10 for LibraryThing, and greater than 0 for HotRec 2011 LastFM. In effect, the latter dataset represents implicit feedback, while the others are examples of explicit feedback. 
Exploiting the lists of the top-10 recommendations for each user, we computed the measures of precision, recall, nDCG, Entropy Based Novelty (EBN)~\cite{bellogin_study_2010}, and diversity~\cite{Zhang:2008:AMI:1454008.1454030}.

For the implementation, we rely on the LibRec library\footnote{\url{https://www.librec.net}}. It computes measures according to the \emph{all unrated items} protocol \cite{Steck2013}. More specifically, it creates a top-N recommendation list for each user by predicting a score for every item not rated by that particular user, whether that item appears in the user test set or not. All the non-rated items are considered to be irrelevant for the user. This explains the low values for the measures (e.\,g., precision and recall) as the quality of recommendations tend to be underestimated. However, Steck \cite{Steck2013} suggests to rely on this protocol rather than the \emph{rated test-items}, which includes only rated test items in the top-N list, as the user satisfaction regarding top-N recommendations depends on the ranking of all items.

\section{Evaluation Results}\label{sec:res}
We report the results of the first experiment on optimizing the parameters of our SemRevRec system in \Cref{sec:exp1}.
The results of comparing our approach with baselines from related work are documented in \Cref{sec:exp2}.

\subsection{Optimizing the SemRevRec Parameters}\label{sec:exp1}
In this experiment, we evaluated the impact of the ranking function, the discovery, the occurrence threshold, and the coefficients of R3 on the performance of our algorithm. We executed SemRevRec in three domains with different ranking functions and with and without the discovery phase. We also varied the configuration parameters $\eta$ and $\kappa$ of the ranking function R3, in order to identify possible relationships between the occurrence and the LDSD measure. Moreover, we considered how the percentage of the minimum occurrence required for entities to be included in the candidate recommendation set impacts on the results. The main configurations tested are listed in Table~\ref{table:config}.

\begin{table}
\centering
\caption{Configuration of SemRevRec}
\label{table:config}
\begin{tabular}{@{}llllll@{}}
\toprule
Conf. & Ranking & Discovered & Occurrence & $\eta$ & $\kappa$ \\ \midrule
C1    & R1      & False      & 0.05       & --     & --       \\
C2    & R1      & True       & 0.05       & --     & --       \\
C3    & R2      & False      & 0.05       & --     & --       \\
C4    & R2      & True       & 0.05       & --     & --       \\
C5    & R3      & False      & 0.05       & 0.50   & 0.50     \\
C6    & R3      & True       & 0.05       & 0.50   & 0.50     \\
C7    & R3      & True       & 0.05       & 0.75   & 0.25     \\
C8    & R3      & True       & 0.05       & 0.25   & 0.75     \\
\bottomrule 
\end{tabular}
\end{table}

Table~\ref{table:ex1-ml}, Table~\ref{table:ex1-lt}, and Table~\ref{table:ex1-fm} summarize the results obtained with the DBpedia knowledge base in the movie, book, and music domain, respectively. For all the measures but EBN, higher values mean better results, while the lower is EBN, the higher is the novelty. The best values and configurations are highlighted with a bold font.\footnote{More values are highlighted for the same measure if the differences among them are not statistically significant.} In order to decide if the difference between two measures was statistically significant, we relied on the Welch's \emph{t}-test (or unequal variances \emph{t}-test), which is an adaptation of the Student's \emph{t}-test more reliable when the two samples have unequal variances and unequal sample sizes~\cite{RePEc:oup:beheco:v:17:y:2006:i:4:p:688-690}. We considered \emph{p}~<~0.001 because we applied the Bonferroni correction as we performed pairwise comparisons.

The obtained results suggest that the discovery of additional entities through Linked Data is useful for improving the precision of the recommended items. 
In fact, the best configurations in all the domains but music (C8 for movies, C2 for books) rely on it. In the music domain there is not a significant difference in the measures when relying on the discovery phase. This may be related to the fact that we considered reviews about musical works in order to recommend musical artists.

The best ranking function depends instead on the domain. For movies, R3 outperformed the other rankers (C8), while, for book and music recommendations, R1 accounts for the best results (C2), although in the music domain the values obtained with R1 and R2 were equivalent (C4). This suggests that a simpler ranker may be more effective on sparse data, and it could be better to rely on information from reviews than on Linked Data. Additionally, the coefficients $\eta$ and $\kappa$ of R3 may have a high impact on the results as shown by C6, C7, and C8 in Table~\ref{table:ex1-ml}, even if, in the music domain, the measures do not vary. In particular, C8 improves significantly the precision and recall measures with respect to other configurations of R3 in the movie and book domains.

\begin{table}
\centering
\caption{Results with MovieLens and DBpedia}
\label{table:ex1-ml}
\begin{tabular}{@{}lrrrrr@{}}
\toprule
Conf.       & Precis.         & Recall          & nDCG            & EBN             & Divers. \\ \midrule
C1          & 0.0604          & 0.0399          & 0.0412          & 1.2804          & \textbf{0.2431} \\
C2          & 0.0529          & 0.0327          & 0.0343          & 1.2776          & 0.1629  \\
C3          & 0.0604          & 0.0399          & 0.0412          & 1.2804          & \textbf{0.2431} \\
C4          & 0.0276          & 0.0178          & 0.0197          & \textbf{0.7820} & 0.1716  \\
C5          & \textbf{0.0683} & 0.0424          & 0.0491          & 1.0047          & 0.1795  \\
C6          & 0.0460          & 0.0255          & 0.0320          & 0.9354          & 0.1794  \\
C7          & 0.0344          & 0.0191          & 0.0243          & 0.8248          & 0.1464  \\
\textbf{C8} & \textbf{0.0711} & \textbf{0.0478} & \textbf{0.0524} & 1.0163          & 0.2114  \\ \bottomrule
\end{tabular}
\end{table}

\begin{table}
\centering
\caption{Results with LibraryThing and DBpedia}
\label{table:ex1-lt}
\begin{tabular}{@{}lrrrrr@{}}
\toprule
Conf.       & Precis.         & Recall          & nDCG            & EBN             & Divers. \\ \midrule
C1          & 0.0396          & 0.0350          & 0.0341          & 0.4081          & 0.7701  \\
\textbf{C2} & \textbf{0.0506} & \textbf{0.0497} & \textbf{0.0465} & 0.2771          & 0.7780  \\
C3          & 0.0396          & 0.0350          & 0.0341          & 0.4081          & 0.7701  \\
C4          & 0.0357          & 0.0340          & 0.0353          & \textbf{0.1946} & 0.8919  \\
C5          & 0.0462          & 0.0373          & \textbf{0.0462} & 0.2809          & 0.8663  \\
C6          & 0.0356          & 0.0331          & 0.0366          & 0.2280          & 0.9039  \\
C7          & 0.0306          & 0.0269          & 0.0317          & 0.2444          & 0.8932  \\
C8          & 0.0421          & 0.0418          & \textbf{0.0429} & 0.2077          & \textbf{0.9118} \\ \bottomrule
\end{tabular}
\end{table}

\begin{table}
\centering
\caption{Results with LastFM and DBpedia}
\label{table:ex1-fm}
\begin{tabular}{@{}lrrrrr@{}}
\toprule
Conf.       & Precis.         & Recall          & nDCG            & EBN             & Divers. \\ \midrule
\textbf{C1} & \textbf{0.0495} & \textbf{0.0504} & \textbf{0.0486} & 0.7894          & 0.5654  \\
\textbf{C2} & \textbf{0.0504} & \textbf{0.0515} & \textbf{0.0473} & 0.6640          & 0.6021  \\
\textbf{C3} & \textbf{0.0495} & \textbf{0.0504} & \textbf{0.0486} & 0.7894          & 0.5654  \\
\textbf{C4} & \textbf{0.0504} & \textbf{0.0515} & \textbf{0.0473} & 0.6640          & 0.6022  \\
C5          & 0.0363          & 0.0371          & 0.0378          & 0.2619          & 0.9238  \\
C6          & 0.0360          & 0.0370          & 0.0378          & \textbf{0.2422} & \textbf{0.9325}  \\
C7          & 0.0361          & 0.0369          & 0.0378          & \textbf{0.2425} & \textbf{0.9325}  \\
C8          & 0.0360          & 0.0368          & 0.0378          & \textbf{0.2411} & \textbf{0.9329} \\ \bottomrule
\end{tabular}
\end{table}

\begin{figure}
\centering
\includegraphics[width=0.46\textwidth]{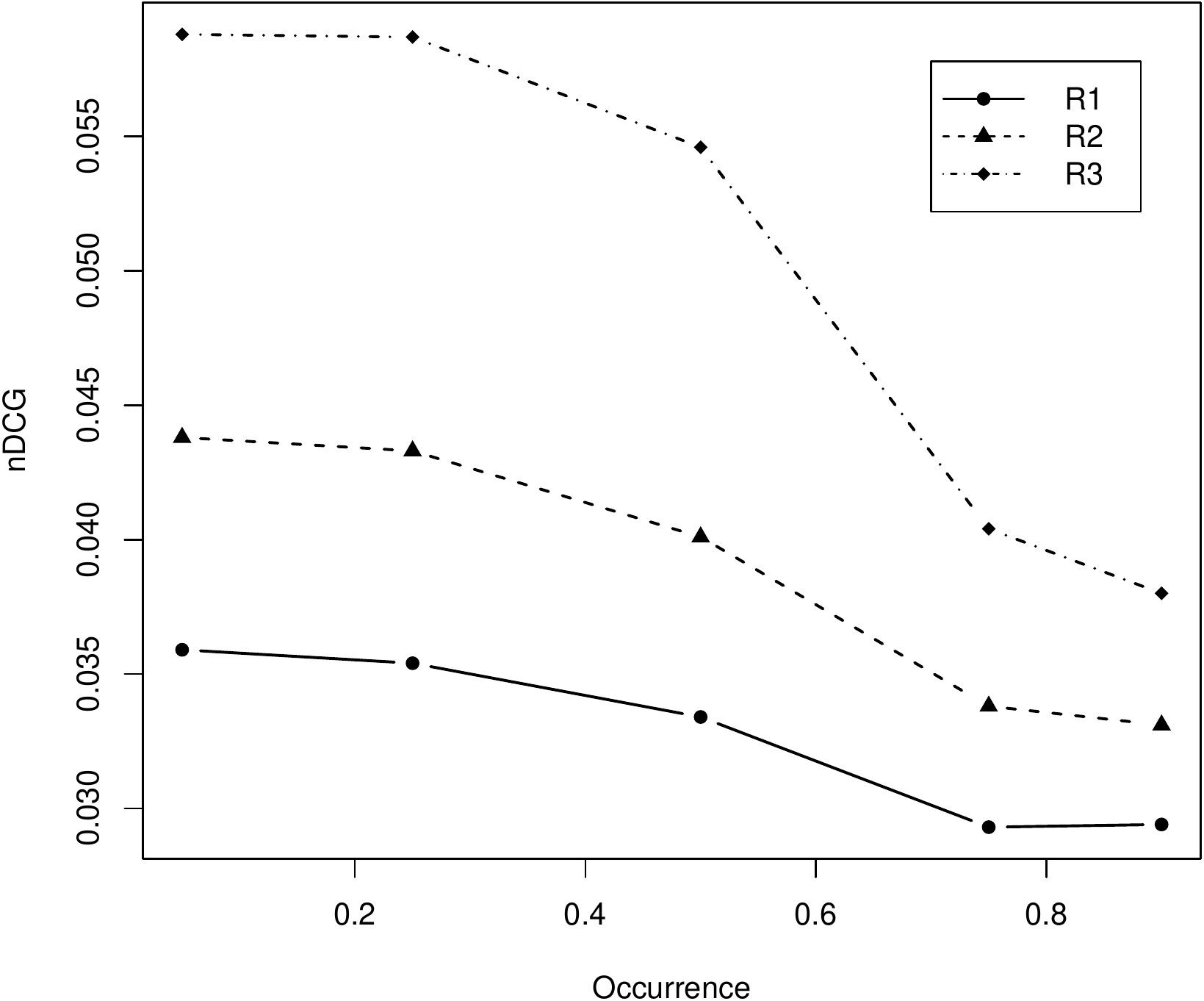}
\caption{nDCG with MovieLens and Wikidata}
\label{fig:lineplot}
\end{figure}
Figure~\ref{fig:lineplot} illustrates the performance in terms of nDCG of the three ranking functions available in SemRevRec when the number of entities considered for the recommendation process varies. The occurrence represents the minimum number of times an entity needs to be annotated in the reviews of a certain item in order to be included in the candidate recommendation set. It is expressed as a percentage of the most annotated entity for an item. The plot is based on the results obtained in the movie domain with the Wikidata knowledge base, as this can be considered the most representative case. Unsurprisingly, all rankers tend to converge, as the number of entities available decreases. However, it is important to notice that the nDCG is monotonically decreasing. This fact happens in the majority of the domains with both knowledge bases and supports the hypothesis that the higher is the number of available entities, the better is the quality of the recommendations.

\begin{figure*}
\centering
\includegraphics[width=\textwidth]{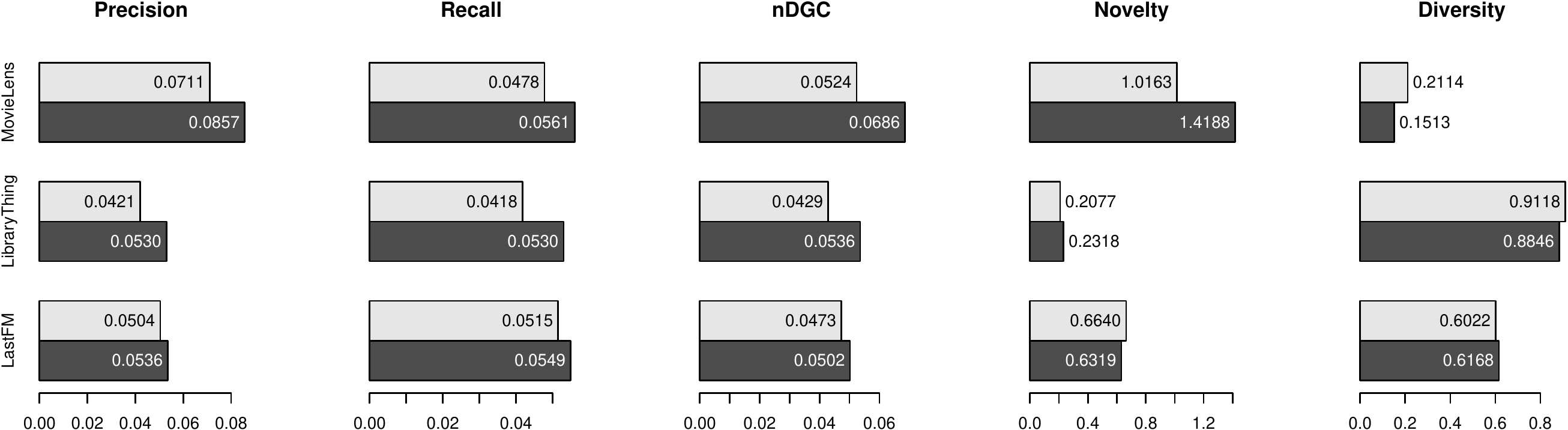}
\caption{Comparison between DBpedia and Wikidata. Light grey represents DBpedia, dark grey Wikidata.}
\label{fig:barplot}
\end{figure*}

Figure~\ref{fig:barplot} compares the results obtained by the best configuration of our algorithm when using DBpedia and Wikidata for each domain. Although both knowledge bases are derived from Wikipedia, the results differ. In particular, Wikidata outperformed DBpedia in the vast majority of the considered measures. A possible reason may be that Wikidata provides higher data quality for the recommendation task, as it also contains knowledge manually encoded by human editors. At the instance level, this may be primary due to  the interlinking of resources since we rely on the LDSD measure which exploit direct and indirect links. At the ontology level, the properties considered in the discovery may also have an high impact. We should investigate which features of a knowledge base are well suited for a Linked Data based recommender system, although they can also depend on the particular domain considered.

Table~\ref{table:ex1-wd} lists the results obtained with Wikidata. They vary significantly when the $\eta$ and $\kappa$ weights of the ranking function R3 are changed. Thus, we decided to include in this paper only the results related to the configurations C4, C6, C7, and C8, although we tested all the ones listed in Table~\ref{table:config}. The complete evaluation is available on the Web.\footnote{\url{https://doi.org/10.6084/m9.figshare.5074081}} In general, Wikidata provides better results with respect to DBpedia and this behavior is consistent in all domains, but differences are more significant when movies are recommended.

\begin{table}
\centering
\caption{Results with Wikidata}
\label{table:ex1-wd}
\begin{tabular}{@{}llrrrrr@{}}
\toprule
Conf. & Domain  & Precis. & Recall & nDCG   & EBN    & Divers. \\ \midrule
C4    & Movie   & 0.0582  & 0.0368 & 0.0438 & \textbf{1.3626} & 0.1223 \\
C6    & Movie   & 0.0757  & 0.0487 & 0.0588 & 1.4284 & \textbf{0.1461} \\
C7    & Movie   & 0.0728  & 0.0459 & 0.0552 & 1.4322 & \textbf{0.1423} \\
\textbf{C8} & \textbf{Movie} & \textbf{0.0857} & \textbf{0.0561} & \textbf{0.0686} & 1.4188 & \textbf{0.1513} \\
\midrule
C4    & Book   & 0.0392   & 0.0373 & 0.0379 & 0.2634 & 0.8455  \\
C6    & Book   & 0.0452   & 0.0443 & 0.0466 & 0.2621 & 0.8705  \\
C7    & Book   & 0.0365   & 0.0334 & 0.0380 & 0.2809 & 0.8600  \\
\textbf{C8} & \textbf{Book}  & \textbf{0.0530} & \textbf{0.0530} & \textbf{0.0536} & \textbf{0.2318} & \textbf{0.8846} \\
\midrule
\textbf{C4} & \textbf{Music} & \textbf{0.0536} & \textbf{0.0549} & \textbf{0.0502} & 0.6319 & 0.6168 \\
C6    & Music  & 0.0384   & 0.0395 & 0.0375 & \textbf{0.3083} & \textbf{0.9314} \\
C7    & Music  & 0.0390   & 0.0401 & 0.0380 & \textbf{0.3062} & \textbf{0.9327} \\
C8    & Music  & 0.0367   & 0.0377 & 0.0363 & \textbf{0.3178} & \textbf{0.9322} \\ \bottomrule
\end{tabular}
\end{table}

\subsection{Comparison with Baselines}\label{sec:exp2}
We compared our technique to the Most Popular, Random Guess, Item KNN, and Bayesian Personalized Ranking (BPR)~\cite{rendle_bpr_2009} algorithms, as implemented in LibRec, and with SPrank~\cite{noia_sprank_2016}, a state-of-the-art Linked Data-based recommender. We set the neighborhood size for Item KNN to 80, while we used 100 factors for BPR, as done by Musto et al.~\cite{Musto:2016:SGR:2930238.2930249}. We configured SPrank to exploit LambdaMart as the ranking method and to follow in the DBpedia graph the same properties that we selected for our algorithm, as listed in Table~\ref{table:disc}.

Table~\ref{table:ex2-ml}, Table~\ref{table:ex2-lt}, and Table~\ref{table:ex2-fm} list the results obtained in the movie, book, and music domain, respectively. The best values are highlighted with a bold font.\footnote{More values are highlighted for the same measure if the differences among them are not statistically significant. In the case of EBN and diversity, when Random Guess was the best, we also highlighted the second best because its precision, recall, and nDCG were close to zero. This means that the recommendations provided are completely unrelated and their novelty and diversity is not relevant.} For SemRevRec, we reported both the configuration with the best trade-off among the various measures and the best scores achieved for each measure in the experiment described in Section \ref{sec:exp1}.
In all the experimental trails, SemRevRec provided the best diversity and a better accuracy (both in rating prediction and ranking) than SPrank, while it improved in novelty with respect to traditional techniques. BPR accounted for the highest precision, recall, and nDCG. In general the diversity of algorithms is rather low for movies, while for music and books is above 0.6, apart for Item KNN.

The differences between SemRevRec and the other approaches are statistically significant according to the Welch's \emph{t}-test with \emph{p}~<~0.001, except for SPrank, BRP, Most Popular, and Random Guess in the movie domain regarding the measure of diversity, SPrank in the book domain regarding the measures of precision and diversity, and Most Popular in the music domain regarding the measure of diversity.

\section{Discussion}\label{sec:discussion}
In general, the results obtained by our algorithm in the music and book domains are not as good as the ones reached with movie recommendations. This may be due to the characteristics of the reviews, as illustrated in Figure~\ref{fig:boxplot} and previously discussed. The entities annotated for each item in these two domains are much less than the entities available in movie reviews. This fact should be further studied. Moreover, it would be interesting investigating the impact of the number of reviews available and their quality with respect to the recommendation process. For example, a meaningful album review mentions the author and similar albums or artists the user liked, while a review describing the package is not very useful in our scenario. In fact, we aim to suggest other artists to listen to, although packaging may impact on the decision of buying a physical copy of that album. Finally, the significant difference in the results obtained when exploiting Wikidata or DBpedia suggests that the impact of knowledge bases, notably the selection of types and properties exploited, on the performance should be further analyzed.

\begin{table}
\centering
\caption{Comparison using the MovieLens dataset}
\label{table:ex2-ml}
\begin{tabular}{@{}lrrrrr@{}}
\toprule
Algorithm & Precis. & Recall & nDCG   & EBN    & Divers. \\ \midrule
SemRevRec & 0.0857  & 0.0561 & 0.0686 & 1.4188 & 0.1513 \\
-- Best Scores & 0.0857 & 0.0561 & 0.0686 & \textbf{0.7820} & \textbf{0.2431} \\ \midrule
SPrank    & 0.0445  & 0.0254 & 0.0280 & 0.8813 & 0.1612 \\
Item KNN  & 0.1626  & 0.1105 & 0.1302 & 2.6846 & 0.0696 \\
BPR       & \textbf{0.2347}  & \textbf{0.1737} & \textbf{0.1930} & 1.8358 & 0.1769 \\
Popular   & 0.1325  & 0.0840 & 0.0969 & 2.7439 & 0.1412 \\
Random    & 0.0055  & 0.0028 & 0.0031 & \textbf{0.3018} & 0.1679 \\ \bottomrule
\end{tabular}
\end{table}

\begin{table}
\centering
\caption{Comparison using the LibraryThing dataset}
\label{table:ex2-lt}
\begin{tabular}{@{}lrrrrr@{}}
\toprule
Algorithm & Precis. & Recall & nDCG   & EBN    & Divers. \\ \midrule
SemRevRec & 0.0530  & 0.0530 & 0.0536 & 0.2318 & 0.8846 \\
-- Best Scores & 0.0530  & 0.0530 & 0.0536 & 0.1946 & \textbf{0.9118} \\ \midrule
SPrank    & 0.0379  & 0.0346 & 0.0337 & \textbf{0.1562} & 0.8037 \\
Item KNN  & 0.0620  & 0.0564 & 0.0662 & 1.4956 & 0.2259 \\
BPR       & \textbf{0.0862}  & \textbf{0.0817} & \textbf{0.0895} & 0.6043 & 0.7177 \\
Popular   & 0.0423  & 0.0343 & 0.0447 & 1.6034 & 0.6483 \\
Random    & 0.0004  & 0.0002 & 0.0003 & \textbf{0.0382} & \textbf{0.9879} \\ \bottomrule
\end{tabular}
\end{table}

\begin{table}
\centering
\caption{Comparison using the LastFM dataset}
\label{table:ex2-fm}
\begin{tabular}{@{}lrrrrr@{}}
\toprule
Algorithm & Precis. & Recall & nDCG   & EBN    & Divers. \\ \midrule
SemRevRec & 0.0536  & 0.0549 & 0.0502 & 0.6319 & 0.6168 \\
-- Best Scores & 0.0536  & 0.0549 & 0.0502 & 0.2411 & \textbf{0.9329} \\ \midrule
SPrank    & 0.0156  & 0.0158 & 0.0176 & \textbf{0.1834} & 0.9077 \\
Item KNN  & 0.1392  & 0.1428 & 0.1720 & 1.6023 & 0.4730 \\
BPR       & \textbf{0.1545}  & \textbf{0.1583} & \textbf{0.1808} & 0.9404 & 0.6547 \\
Popular   & 0.0686  & 0.0703 & 0.0791 & 2.0360 & 0.6519 \\
Random    & 0.0005  & 0.0005 & 0.0004 & \textbf{0.0442} & \textbf{0.9946} \\ \bottomrule
\end{tabular}
\end{table}

In this work, we relied on all the reviews available for the items present in the rating datasets used for the evaluation. However, only reviews about some items, i.\,e. the ones with the average rating higher than a threshold, or only some reviews for each item, i.\,e. only the ones which are rated positively, could be considered during the semantic annotation phase. Nevertheless, lower performance on music artists and books was expected because the available ratings were more sparse than the ones regarding movies. This holds for all the algorithms and explains the general difference of scores in these domains (overall lower than for movies). 

SemRevRec showed the best diversity in all the domains. Notably, in the sparse dataset of books, it achieved precision, recall, and nDCG comparable to Item KNN with a much higher diversity, although both are content based methods. However, collaborative filtering techniques are know to suffer less of the overspecilization problem and provide better rating prediction and ranking than content based ones as SemRevRec. For this reason, although collaborative filtering is very popular, we decided to include in the baseline only one technique among many, i.\,e. BPR, which is one of the newest and most promising. Nevertheless, it showed a lower diversity than our algorithm. Not surprinsigly, it also accounted for the best rating prediction and ranking. 

Our approach also provided a higher novelty than traditional techniques and a better rating prediction and ranking than SPrank. In the movie domain, SemRevRec accounted for the best novelty, while with music and books for the second best, with results close to SPrank. Additionally, when optimized for this measure, SemRevRec had similar (for books) or higher (for music) rating prediction and ranking than SPrank. On the contrary, when the former is optimized for rating prediction and ranking, it could be preferred to the latter to increase the novelty of recommendations, but also limiting the loss in rating prediction and ranking. Additionally, SemRevRec was evaluated considering the recommendations generated for all the previous movies a user liked since its generation approach is rather naive and takes into account only an initial item. Combining it with a machine learning technique could significantly improve its performance, but further experiments are required to prove this.
\section{Conclusions and Future Work}\label{sec:concl}
In this paper we proposed a novel recommendation approach, based on the semantic annotation of user reviews and Linked Data. We conducted an offline study of the recommender system in the movie, book, and music domains, which showed that our method provides the best diversity. It also improved rating prediction and ranking compared to another algorithm based on Linked Data, while it increased the novelty of recommendations with respect to traditional techniques. We also tested our approach with different knowledge bases and Wikidata systematically achieved better results than DBpedia.
Although the reviews available for the book and music domains seem to contain a smaller amount of useful information, the results of the offline study suggest that our algorithm can provide more diverse recommendations and reach an interesting compromise between the accuracy and the novelty of the suggested items.

As future work, we intend to investigate in greater details how the nature of the user reviews influences the performance of our algorithm. Moreover, the significant difference in the results obtained when exploiting Wikidata or DBpedia suggests that too little is known about how knowledge bases (notably their types and properties) might impact on the performance of Linked Data based recommender systems. We also plan to take into account also the sentiment of the reviews, i.\,e. whether the overall opinion on the item reviewed is positive or negative. Finally, we are evaluating applications of our approach on textual resources different than reviews, e.\,g. research papers or their abstracts. In this case sentiment would not be relevant, while annotated entities could be concepts representing the main topics addressed in the document. This suits recommendation of research papers.

\begin{acks}
This work was supported by the EU's Horizon 2020 programme under grant agreement H2020-693092 MOVING.
\end{acks}

\bibliographystyle{ACM-Reference-Format}
\bibliography{ms} 

\end{document}